# Where Backtests and Live Trading Part Ways: Six Defects from One Intraday System

Alex Vidovich[a,*], Zivana Zerjal[b]

[a] *Independent researcher · alexvidovich.com*

[b] *Independent researcher · zivanazerjal.com*

** Corresponding author.*
*E-mail addresses:* *support@alexvidovich.com (A. Vidovich), support@zivanazerjal.com (Z. Zerjal)*

A B S T R A C T

Backtests of intraday strategies can drift away from what the live system actually does; in our case, the drift never showed in summary statistics. We report six defects found while running one intraday US equity system live and comparing its live days with their backtest replay. Three came from data: a default feed that carried a small fraction of real opening volume, historical prices adjusted for dividends that live orders never see, and data the live system downloaded but the backtest never did. Two came from code: a sizing step that exceeded a limit the live account enforces, and UTC timestamps read as New York time. One came from configuration: settings that were switched on but never reached the code that places orders. For each we give the symptom, the size of the error, the cause and the test or control that now guards against it. All six sat where two paths that should agree had drifted apart. We close with a ten-point checklist. We make no claim about whether the system is profitable.

Figures come from the authors' dated research log, which is not public. Some figures are rounded or withheld to protect the live system.

## 1. Introduction and setting

**Related work.** Most writing on backtest failure is about statistics: overfitting from many trials (Bailey, Borwein, López de Prado & Zhu 2014; López de Prado 2018) and research protocols that control it (Arnott, Harvey & Markowitz 2019). Data problems in high-frequency market data are documented for academic datasets (Holden & Jacobsen 2014). Configuration and pipeline debt is well described for machine-learning systems (Sculley et al. 2015). This paper sits between these. It reports implementation defects in data, code, clocks and configuration. They bias a backtest before any statistics run. The safeguards in that literature assume correct inputs, so they cannot catch these.

**The system.** The system trades a fixed universe of 97–98 US stocks intraday from 1-minute bars, and makes its decision shortly after the open. It has run in production on a paper account and with real money. Its entry, exit, filter and sizing rules are not described here. None of the defects needs them.

**The research stack.** Market data comes from Alpaca's market-data API. One fetch function serves both the backtest cache builder and the live runner. Bars are cached to disk by date, and research scripts read the cache. We call the main backtest the **parity harness**. It drives the same strategy code the live runner uses, with its daily inputs and decision-time quotes fetched from the broker, and it is checked against live fills. **Research scans** are one-off scripts that read the cache directly.

**Parity with live is the first test a harness must pass.** A backtest is a claim about what the live system would have done, so we replay days the live system actually traded and compare. We check both paper and real-money days. The difference matters: a paper account's fills are simulated by the broker, so a paper-fill match tests the harness against the broker's simulator, not against the market. Real-money days count for more. Most harness figures in this paper, including every parity check below, were produced while Defect 5 was present. The text says where its effect is known; elsewhere it is unmeasured.

- *Real-money days.* Five real-money days were replayed. On two, an execution check that exists only in live changed what was traded; such trades are excluded from the slippage test by a rule set in advance. Two matched live and passed a pre-registered slippage bar: replay and live results for the same trade may differ by 5% of the replay result or by a small fixed dollar floor, whichever is larger. The fifth day diverged. It was later traced to Defect 5; with the fix, the replay takes live's trade and passes the bar. A later real-money pass turned out to be two errors cancelling (Section 3); with Defect 5 fixed and that second gap still open, the same day now fails the bar on a sizing artifact. The two passes above were unchanged by the fix.
- *Earlier, mostly paper days.* After the fixes for Defects 1 and 2, the harness matched the live decision on 5 of 5 days, a result produced while Defect 5 was present and so provisional. Four were paper fills and one was real money; that day predates the slippage bar. At the account's real size limit it reproduced live share counts and results within slippage.

The live-only execution checks are deliberate: they need live data the harness does not replay. Section 5 lists them as a known gap.

Per-trade dollars differ even on matched days, because the harness uses a cached spread and live uses the real quote. So we never trust a single-trade backtest figure on its own. We trust the harness's decisions and aggregates more than its single trades, and only as far as the evidence above reaches. That trust holds only once the inputs are right, and Defect 5 shows how long a wrong input can go unseen. Parity shows that the harness copies live. It does not show that its results generalize. Every parity day so far fell in US summer time (Section 5).

A second harness in the same repository shows what happens without the parity test. It derived its daily inputs offline instead of fetching them. At the same size limit it returned less than half the parity harness's figure, because it misread its inputs and dropped the trades that moved the total most. It told a confident, wrong story about which days drove the result. It was caught when one of the authors remembered a different shape and asked for a check, and it was retired the same day. That same retired harness had matched the live decision on 7 of 8 real-money dates, and the later correction in our log does not dispute that match. So matching decisions on the days checked did not catch a broken harness. Matching decisions is necessary, not sufficient.

## 2. The six defects

Each defect has four parts. **Symptom** is what we saw. **Magnitude** is how much the answer moved, as the log states it. **Root cause** is what was wrong. **Detection test** is the automated test, or where none exists the control, that now guards against it. Tests are described rather than named. Appendix B gives one worked example.

### *2.1 Defect 1 — The default feed is not the consolidated tape*

**Symptom.** While preparing a six-month A/B backtest, we compared one live trade against the cached bars for the same minutes. The cache's opening-minute volume was a tiny fraction of what live had seen. A scan of all 98 cached 2026 trading days found at least one liquid name with a 9:30 bar under 50,000 shares on every single day.

Two details hid it. Full-day volume totals in the cache were correct; only the per-minute split at the open was thin. And live was unaffected: its decision-time fetch matched the consolidated tape to the share.

**Magnitude.** Two log entries disagree on the range: 0.1–0.5% of true opening-minute volume and 0.1–2%. We use the later, wider range.

**Table 1. Opening-minute volume, same minute, two sources** (stocks anonymized; shares rounded)

| Comparison | Minute (ET) | Shares | Share of consolidated |
|---|---|---|---|
| Consolidated tape (SIP), Stock A | 9:30 | ~650,000 | 100% |
| Single-exchange feed (IEX), same minute | 9:30 | ~11,000 | — |
| No feed parameter, same minute | 9:30 | same as IEX | — |
| Cache vs consolidated, Stock B | 9:30 | ~7,900 vs ~3.4 million | 0.2% |
| Cache vs consolidated, Stock A | 9:30 | ~2,600 vs ~550,000 | 0.5% |
| Cache vs consolidated, Stock A | 9:32 | ~3,000 vs ~210,000 | 1.4% |

Every feature built from opening-minute data was wrong. For one stock and minute, the difference was large enough to flip a decision. Opening prices were off as well.

On two months replayed both ways, the same trades were chosen. The thin cache understated one month and overstated the other, and in the second it turned one losing trade into a winning one. Two months, one error each way, cannot separate noise from bias. What they do show is that one month's figure could be off in either direction with nothing in the summary to say so. This comparison was made while Defect 5 was present; whether Defect 5 touched it is unmeasured.

The defect also voided an earlier finding. A permutation test run on the thin cache (p = 0.217) was marked invalid and later retired.

**Root cause.** The fetch request carried no feed parameter. In our test, a historical request without one returned single-exchange (IEX) data instead of the consolidated tape (SIP); Table 1's third row is the proof. Alpaca documents the default as the best feed the user's subscription allows (Alpaca 2026). We did not establish why the live runner's same-day requests, through the same function, matched the consolidated tape. The mistake was ours: we never pinned the parameter. While checking the fix we also hit the clock trap that later became Defect 4.

**Detection test.** Two layers.

- A request-recording test calls the fetch function through a fake session that records every request, and asserts that every request pins the consolidated feed. Appendix B shows the pattern.
- A runtime check covers what a pinned parameter cannot: a new code path or a vendor change. At decision time, if opening volume has collapsed against the prior day across nearly the whole universe at once, the live system logs a warning and an alert event. It never blocks a trade. Its unit test feeds in a universe-wide collapse and asserts that every name is flagged. The reasoning, stated in the test: a collapse like that across the whole universe at once is not a normal market event. A market-wide disrupted open could trip it, and its false-alarm rate is unknown (Section 5).

**Lesson.** Pin every vendor parameter that changes what the data means, even when today's default looks right. Then check, at the point of use, a property of the data the bug would break.

### *2.2 Defect 2 — Dividend-adjusted history against unadjusted live fills*

**Symptom.** After Defect 1 we rebuilt the dataset and ran the parity harness. One replayed trade, a paper-account fill, did not match. The price-basis problem does not depend on who fills the order. One bar in the rebuilt dataset sat 0.68% below the live price basis, with volume identical to the share. A same-day scan found 37 of the 98 symbols shifted the same way, every one a dividend payer. The log's examples range from −0.68% to −1.65%.

**Magnitude.** On that trade, the harness combined an unadjusted entry price with an adjusted exit price. The replay result came out far from the live result on the same trade. The first two full runs on the new dataset were invalidated.

**Table 2. One decision-time bar, Stock C, four sources**

| Source | Close vs live fill basis | Volume | Price basis |
|---|---|---|---|
| Live order | reference | — | unadjusted |
| Old cache | matches | identical | unadjusted |
| Fresh pull, no adjustment setting | matches | identical | unadjusted |
| Rebuilt dataset, adjustment "all" | −0.68% | identical | adjusted |

After re-pulling unadjusted prices, entry and exit on that trade matched live to the cent. The remaining gap was share count only, a known sizing difference.

**Root cause.** Three and a half weeks before discovery, a commit switched the fetch's adjustment setting from split-only to "all", which on Alpaca includes dividends. That setting restates a historical bar on the latest price basis. A bar from date D, fetched later, has every dividend after D subtracted. Live fetches same-day bars, when no later dividend exists, and live orders fill at unadjusted prices anyway. So the setting was harmless in live and wrong in every backtest.

**Detection test.** A request-recording test, the same pattern as Defect 1's, asserts that every request asks for unadjusted ("raw") prices. What found the defect was not a unit test. It was the parity step: before trusting a rebuilt dataset, replay one live trade and match entry and exit to the real fill.

**Lesson.** Adjusted prices answer "what was this worth, in today's terms". A backtest must answer "what price would my order have filled at that day". Use unadjusted prices for execution simulation.

### *2.3 Defect 3 — A sizing step that scaled past the binding limit*

**Symptom.** A live loss prompted a line-by-line read of that day's raw decision record. Two size numbers disagreed: a per-trade ceiling, and the size live actually took, which the account's buying power capped. Live was right: buying power was the real limit, and it binds well below that ceiling. Tracing the same arithmetic through the backtest found the reverse problem. There, a later sizing step could push a position above the limit.

**Magnitude.** In the reference backtest, 20 of 61 simulated fills were sized at a notional the live account can never take. The log estimated that backtest's dollar figures at about 18% too large in absolute size, and ordered every earlier dollar figure discarded. The fix changed notionals, not which trades were taken: the fill count was 61 before and after. This backtest was run while Defect 5 was present, so Defect 5's effect on these figures is unmeasured.

It was the second defect of this shape. Nine days earlier the harness had not modeled buying power at all and oversized the real account several-fold; Defect 3 is the part of that gap the first fix left behind. A third instance appeared later (Section 3).

**Root cause.** Sizing is a chain of steps. The strategy code applied the buying-power limit correctly. A later step, one that can increase size, then multiplied the already-limited notional with no second limit. Live was never exposed: it re-checks real buying power at order time, and the field the late step read exists only in the backtest.

**Detection test.** A unit test builds a setup already at its limit, applies the size-increasing step, and asserts that the ceiling, the notional and the share count are all unchanged. The fix is structural: each setup now carries its limit, so any later step can re-apply it.

**Lesson.** In a chain of sizing steps, apply the binding limit last, or re-apply it after every step that can increase size. Test with a position already at the limit. That is the case that breaks.

### *2.4 Defect 4 — The clock is UTC, the market is New York*

**Symptom.** While preparing a new study, we noticed a research scan picking the opening bar with the fixed key "13:30". The caches store bar timestamps in UTC. 13:30 UTC is 9:30 ET only during daylight saving time. In winter it is 8:30 ET, an hour before the open.

**Magnitude.** The study period had 129 winter days, 30% of the sample. On a 13-day winter sample, 16% of stock-mornings (one stock on one morning) were scored on a pre-market bar and 84% were silently dropped by a completeness check. The log applies these shares to all winter days. So the scans described about 70% of the period, plus a pre-market slice, and reported it as the whole period. Eleven scans were affected. All were patched and re-run the same morning. A second form of the bug turned up during the re-runs: four scripts built the hour as `13 + minutes // 60`, which hard-codes the summer offset. A run that fixed only the first form briefly printed a positive result for one idea. It vanished with the second fix and was never quoted.

**Table 3. Four rows from the corrected table** (rates omitted)

| Study | Before (UTC key) | After (New York key) | Verdict |
|---|---|---|---|
| A | 2,207 stock-mornings | 3,024 stock-mornings | unchanged |
| B | rejected | rejected | unchanged |
| C | rejected | rejected, all variants negative | unchanged |
| D | count 127 | count 167 | — |

We chose these four of the table's rows to show both outcomes: verdicts that held and numbers that moved. Study A's sample grew by the winter mornings the bug had dropped. In the full table, no verdict flipped, and almost every number moved. A reviewer who checks only verdicts will not see it.

**Root cause.** Clock times were compared as strings, and the string held for one season only. The same trap had been logged months earlier, when matching "09:30" against a UTC timestamp made a correct cache look broken. It was written down as an operational note, not turned into shared code, so it came back. The parity harness and the live runner were never affected. Both parse timestamps into time-zone-aware objects.

**Detection test.** There is no unit test for this yet; we checked while writing this paper. The control is one shared helper that converts cache timestamps to the New

York clock, plus a written rule: a bare string slice of a cache timestamp is a bug on sight. The missing test is simple: one winter date and one summer date, asserting 9:30 ET resolves to the right bar on both.

**Lesson.** Never compare clock times as strings. Convert to the exchange's time zone once, in one shared function. Put a winter day and a summer day in every time-dependent test.

### *2.5 Defect 5 — Data the live path downloads and the replay did not*

**Symptom.** One real-money day would not reconcile: live and the replay made different decisions. The log recorded the cause as an input that differed between the two, and left it open. Three weeks later we chased it. The input depends on data the live system downloads in addition to the stocks it trades. The replay downloaded only the stocks it trades.

**Magnitude.** With that data missing, the input silently read zero in the replay (or was only partly computed, on days with some of the data present), in every harness run for about four months. The input feeds several decision rules, so in the replay they all ran on a wrong value. With the data restored, the replay reproduced live's value exactly and made the same decision as live, within the slippage bar. Across 20 real-money mornings, restoring it changed one decision in twenty. These 20 mornings compared the replay before and after the fix; they were not scored against live.

Two further effects matter more than the one day. First, an earlier difference between live and replay had been put down to different quote timing. The arithmetic of this defect matches that difference exactly; the log calls the re-attribution "likely, not proven". Second, every harness figure produced during those four months carries this defect. Their size and direction of error are unknown, because none has been re-run yet.

**Root cause.** Live fetches its data from its own list of required inputs. The replay built its data request from the list of traded stocks. The two lists had once matched, and drifted apart when the traded list changed. So the research path asked for less data than the live path. And a missing input did not raise an error. It fell back to a neutral default, which here meant zero.

**Detection test.** Three tests were written first and failed for the right reasons. The harness now warns, and records the gap in every day's output, whenever a required input is missing. The harness and the replay now request the live path's own list of required data, and a test asserts that the extra data is in that request. The gap can no longer be silent.

**Lesson.** Build the research data request from the live system's own list of required inputs, never from the list of things it trades. Never let a missing input fall back silently to a neutral value; make it loud. And chase every unexplained parity divergence. This one reached every harness run for four months.

### *2.6 Defect 6 — "Enabled in config" is not "connected in the execution path"*

**Symptom.** This is a family. The log lists six members, and three of them had been named as one class earlier. We show three.

1. *An exit rule that was on and ignored.* The live monitor replays the open trade through the backtest simulator and acts only on exit reasons in an honored set. A new rule's reason was not in the set. Live ignored the rule. Worse, the simulation ended at the ignored exit, which hid every later exit that was honored.
2. *A sizing rule wired only into the backtest.* It was implemented in a branch that runs only when a backtest-only field is set. The live order path never used it, and two live trades were sized as if the rule did not exist. Its "end-to-end confirmation" had checked that the flag loaded into config, not that orders changed. A planned upgrade had been armed into the same dead branch; switching it on would have done nothing.
3. *Invisible settings.* Three settings were read by the code but not declared in the config file, and one of them was on by default. An audit found 53 keys read by production code but undeclared, about a quarter of the surface. While generating those declarations from code defaults, we found the service's startup file carried five stale settings of its own. One, the switch that turns order execution on, existed only there. Shipping the generated default would have silently put the live system into dry-run.

**Magnitude.** No single dollar number. In each case live behavior differed from what the research assumed.

**Root cause.** Settings, code paths and tests each looked right on their own. Nothing checked that a setting changed an observable action. One regression map was itself false assurance: it listed four keys the runtime never reads, so its test passed while guarding nothing.

**Detection test.** Three layers.

- A mapping test reads the declared config and asserts that every exit flag set to on maps to a reason in the live monitor's honored set. A second test pins that every mapped key is really read by production code.

- A presence test asserts that every key the code reads is declared. A neutrality test asserts that loading the declared values gives a config identical to the code defaults, apart from documented exceptions.
- A preflight script, with its own tests, runs twice a day on the live server. It checks behavior in logs and output files, not settings.

**Lesson.** For every setting, test that it changes something observable in the path that places orders. Declare every key the code reads. Then check behavior in production logs, because the config file may not be the only place settings live.

## 3. What the six have in common

**Each sat where two paths that should agree had drifted apart.** Defect 1: one function, a historical request and a live same-day request that came back from different feeds. Defect 2: a historical query and a same-day one. Defect 3: a field that exists only in the backtest. Defect 4: research scans and a harness that parses time correctly. Defect 5: the data list the live path requests and the one the replay requested. Defect 6: configuration and the code that places orders. In the first five, the path closer to live (live itself, or the parity harness) was right. In the sixth, the research modeled behavior live never had. The classes are general; the magnitudes are ours.

**None was found by reviewing summary results.** Defect 1 erred in both directions. Defect 4 moved almost every number and flipped no verdict. Defect 5 changed one real-money decision in twenty. Defect 3 moved every dollar figure, but without a true figure to compare, nothing looked wrong. A bug that makes results absurd gets caught. These made results look plausible.

**Most were found by a narrow, concrete look.** One live trade against the cache. One bar from four sources. One day's decision record, re-read. One script's key, checked before a new study. One unexplained real-money day, chased to the end. The system's own logs, on the first live session after a change. The exception is part of Defect 6: after the third instance of that class, a systematic audit found the undeclared keys. Once a class has recurred, sweeping for it pays.

**Written lessons did not stop recurrence.** The clock trap was logged and came back months later. The sizing-chain lesson of Defect 3 came back as well. Fixing Defect 5 exposed that live and replay applied a size cap at different points in the sizing chain. On one real-money day the two errors cancelled and produced a false pass. We have since turned most lessons into tests, but one data point already says tests alone are not enough: Defect 3's test was in place, and its class came back in a step that test does not cover. The project also keeps a research graveyard, one entry per rejected idea with the test that rejected it; it works like a regression test for ideas. But a rejection is only as good as the harness that made it, so every rejection recorded while Defect 5 was present is provisional until re-run (our reading).

## 4. The checklist

**Table 4. Ten checks, each traced to a defect**

| # | Check | Defect |
|---|---|---|
| 1 | Pin the data feed in every request; test the parameter | 1 |
| 2 | At decision time, alarm if opening volume collapses across the whole universe at once | 1 |
| 3 | Use unadjusted prices for execution simulation; test the parameter | 2 |
| 4 | Replay live trades regularly and match decisions, prices and sizes; after any data change, match one fill to the cent | 2, Sec. 1 |
| 5 | Re-apply the binding size limit after every step that can increase size; test at the limit | 3 |
| 6 | Convert timestamps in one shared exchange-time helper, tested on a winter and a summer day | 4 |
| 7 | Build the replay's data request from the live path's list of required inputs; make any missing input loud, never a silent default | 5 |
| 8 | Chase every unexplained difference between replay and live to its root cause | 5 |
| 9 | For every setting, test that it changes an observable action in the order path; declare every key the code reads | 6 |
| 10 | Check live behavior in logs daily, not settings | 6 |

## 5. Limitations

- **One system, one vendor, one account.** The magnitudes, and the dollar sizes in particular, are ours and do not transfer. Vendor behavior described here is what we observed on our subscription in 2026, and may differ elsewhere or change. The defect classes are general, but we have not shown how often they occur elsewhere.
- **Selection of the six.** We chose one clear instance per class: the one with the clearest record and the clearest test. The paper mentions larger errors that are not among the six (the retired harness in Section 1, the earlier oversizing in Defect 3). An earlier draft carried a method defect: a permutation test run on a population that lacked one of the live system's day-level rules. We dropped it because its deciding evidence was a harness run made while Defect 5 was present. The log does not count its defects.
- **Omitted figures.** To protect the live system, some figures are rounded or left out. The dollar floor of the slippage bar is withheld. On small trades a fixed floor can be looser than 5% of the result, so a reader cannot judge how strict the real-money passes were.

- **Interactions.** The defects were found and fixed one at a time. How they interacted while several were present at once was not measured.
- **Self-report.** The log was written by the same team that made the errors. Entries are dated and most name a commit.
- **Unmeasured effect of Defect 5.** Every harness figure produced while Defect 5 was present carries it, including figures in this paper and rejections recorded in the research graveyard. None has been re-run yet, and the size and direction of the error are unknown.
- **A thin parity basis.** The current harness's real-money record is seven days. One predates the slippage bar. On two, a live-only execution check changed what was traded. Three pass, one of them only after the Defect 5 fix. One passed only because two errors cancelled, and now fails until the size-cap fix is built. Some real-money days have not been replayed on the current harness. The other parity checks used paper fills, which the broker simulates (Section 1). Every parity day fell in US summer time, so no check has yet covered a winter day, which is Defect 4's season.
- **Known gaps between harness and live.** Two execution checks exist only in live, because they need live data the harness does not replay. One size cap is applied at different points in the sizing chain; a fix is queued, not built. The harness also does not model every account-level rule of the live system.
- **Small samples.** Defect 1's dollar effect rests on two months. Defect 4's error shares come from a 13-day winter sample.
- **Controls without tests.** Defect 4 has a shared helper and a rule, but no unit test.
- **Splits.** Unadjusted prices fix dividend drift but leave stock splits in the data. We have not checked how our own features behave across a split date.
- **The volume alarm only warns.** Its false-alarm and miss rates are unknown.
- **Detection bias.** These are the defects a narrow look happened to hit. Others may exist. No test has yet been seen catching a recurrence, and one class has already recurred past its test (Section 3).

## Appendix A. Before and after, per defect

**Table 5. Summary** (figures as stated in the research log)

| Defect | Before the fix | After the fix | How long it lived |
|---|---|---|---|
| 1. Feed | opening-minute volume ~0.1–2% of consolidated | consolidated; matches live to the share | about 11 weeks |
| 2. Adjustment | 37 of 98 symbols shifted; one bar −0.68% | entry and exit match live to the cent | 3.5 weeks |
| 3. Size limit | 20 of 61 fills past the live limit; ~18% too large | all fills within the limit; same 61 fills | start unknown |
| 4. Clock | 129 winter days (30%) mis-read or dropped | all days on New York keys; no verdict flipped | one day in the scans; the trap had been logged months before |
| 5. Missing data | one input read as zero; restoring it changed one decision in 20 replayed real-money mornings | replay requests live's data list; missing inputs flagged | about four months |
| 6. Wiring | exit ignored; sizing rule dead in live; 53 undeclared keys; execution switch in one file only | class-level tests; all keys declared; one config source | exit rule 1 day; sizing rule 10 days live; startup settings since June, removed in July |

## Appendix B. A worked test: record the request, not the response

The tests for Defects 1 and 2 share one pattern. Pseudocode of the pattern (the real tests are separate functions):

```
def test_every_bar_request_pins_feed_and_price_basis():
    recorded = []

    class RecordingSession:                  # stands in for the HTTP client
        def get(self, url, params=None, timeout=None):
            recorded.append(dict(params or {}))
            return EmptyBarsResponse()       # no bars, so no network is needed

    fetch_bars(RecordingSession(), ["SYM1", "SYM2"], start, end)

    assert recorded, "no request was made"
    assert all(p.get("feed") == "sip" for p in recorded)
    assert all(p.get("adjustment") == "raw" for p in recorded)
```

The test checks what the code *asks for*, not what the vendor sends back. It needs no network and no credentials. It fails the moment anyone drops or changes a parameter, and it keeps working even if the vendor's defaults change. The same pattern covers Defect 5: record the symbols the replay requests, and assert they include every symbol on the live path's required-data list.

## Use of AI tools

During the preparation of this work, the authors used Claude (Anthropic) as a research assistant, to help write the project's code and research log, run the logged analyses and draft this text. The authors chose the research questions, made the keep-or-kill decisions and verified the results. The output was reviewed, edited and verified by the authors, who take full responsibility for the content.